\documentclass{vgtc}                          

\usepackage{mathptmx}
\usepackage{graphicx}
\usepackage{times}
\usepackage{algpseudocode}
\usepackage{algorithm}

\onlineid{227}

\vgtcinsertpkg

\title{ Effects of Virtual Acoustics on Dynamic Auditory Distance Perception }

\author{Atul Rungta\
Nicholas Rewkowski\
Roberta Klatzky\
Ming Lin\
Dinesh Manocha}

\teaser{
 \includegraphics[width=7in]{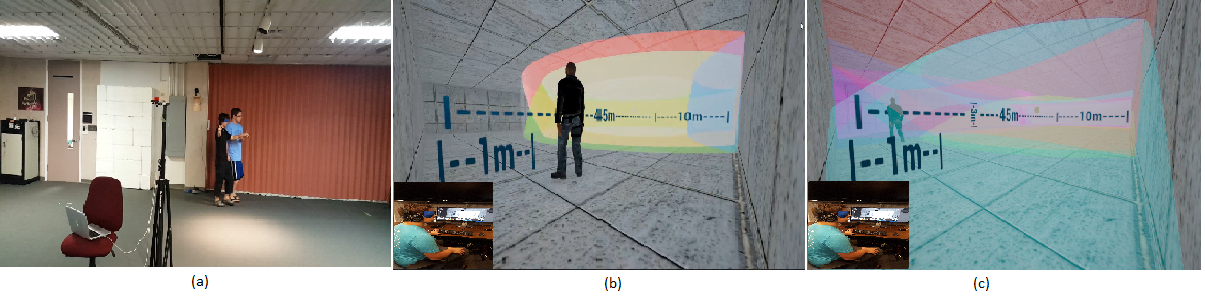}
 \vspace*{-2em}
 \label{img:teaser}
 \caption{Effects of reverberation on acoustic distance perception in dynamic environments. (a) our training setup in a real-world environment with a user being trained; (b) a blindfolded subject taking the study (inset) and the reverberation pattern shown in different colors;  
 (c) the subject moving along a path and experiencing dynamic reverberation. Our study shows that accurate ray-tracing-based reverberation produces a consistently higher perception of distance compared to a commonly-used approximate technique based on a parametric reverberation filter.}
}

\abstract{

Sound propagation encompasses various acoustic phenomena including reverberation. Current virtual acoustic methods, ranging from parametric filters to physically-accurate solvers, can simulate reverberation with varying degrees of fidelity. We investigate the effects of reverberant sounds generated using different propagation algorithms on acoustic distance perception, i.e. how far away humans perceive a sound source.
In particular, we evaluate two classes of methods for real-time sound propagation in dynamic scenes based on parametric filters and ray tracing. Our study shows that the more accurate method shows less distance
compression as compared to the approximate, filter-based method. This 
suggests that accurate reverberation in VR results in a better reproduction
of acoustic distances. We also quantify the levels of distance compression introduced by different propagation methods in a virtual environment.

} 

\begin{document}



\maketitle
\section{Introduction}

Realistic sound effects can increase the sense of presence or immersion in virtual environments. In fact, some  researchers and developers consider spatial audio and sound rendering as VR's second sense\footnote{https://www.youtube.com/watch?v=Na4DYI-WjlI}. Recent hardware developments in virtual reality have resulted in renewed interest in the development of fast techniques to generate plausible or realistic sounds on commodity hardware. 

Sound in virtual environments has received less attention than visual rendering and effects. 
Over the last decade, many new algorithms have been proposed for sound propagation and rendering.
However, there is relatively less investigation on evaluating
the perceptual effectiveness of these methods in virtual environments~\cite{larsson2002better,rungta2016psychoacoustic,zahorik2002assessing}. 


One of the most important acoustic effects is reverberation, which  corresponds to the sound reaching the listener after a large number of successive temporally dense reflections with decaying amplitudes.
Reverberation is known to have multiple perceptual effects on humans, including
auditory distance perception, environment size estimation,
degraded localization, and degraded speech clarity~\cite{Valimaki:2012:FYA:2713201.2713267}. 

Some of the widely used techniques for virtual acoustics are based on 
 artificial reverberation filters. The first set of such algorithms were proposed in the early 1960s, and they have been regularly used for music production, games, and VR. These methods low runtime overhead and most game and VR engines use such filters to generate  propagation or reverberation effects. Recent advancements in computing power and algorithmic methods have resulted in physically-accurate techniques for simulating reverberation at interactive rates based on ray tracing~\cite{schissler2016interactive}, wave-based methods~\cite{Mehra12a}, and ray-wave coupling~\cite{Yero13}. However, some of these accurate techniques~\cite{Mehra12a,Yero13} have a considerably higher precomputation or runtime overhead as compared to artificial filters. 
 An open issue is whether the increase in the physical accuracy of sound propagation leads to an increase in perceptual differentiation.

 In this paper, we evaluate and quantify the relative benefits of different propagation methods for auditory distance perception in VR. In general, distances tend to be under-estimated or under-perceived in virtual environments and this can have considerable impact on the resulting simulation or training applications~\cite{Finnegan:2016:CDC:2858036.2858065}.
 Distance perception has been actively studied with respect to visual and aural cues. In particular, it is known that distance perception in vision, as assessed by direct walking, is accurate to more than $20$ meters~\cite{loomis2003visual}. Auditory distance perception, in contrast to visual cues, is compressive~\cite{loomis2002spatial}. In reverberant environments, there is less compression in terms of auditory distance perception~\cite{klatzky2003encoding}.
 One of the most important distance cues is the loudness of the sound and this cue can be estimated more accurately when the energy ratio of the direct and reverberant sound is available. Some of the earlier work on auditory distance perception was based on setting up physical environments, using appropriate sound sources, and performing the studies. However, prior studies of auditory distance perception have been  affected by the issues that arise in terms of controlling and quantifying the actual sound stimulus to the ears of the users in a physical scene.  In contrast, virtual acoustic techniques can provide more flexibility and less expensive solutions in terms of evaluating auditory distance perception~\cite{bronkhorst1999auditory}. In that regard, the recent developments in interactive and physically-accurate sound propagation methods can provide unique capabilities for performing novel experiments using virtual acoustics.
 
 \nocite{Boustila:2015:EFA:2820619.2820627}

\noindent {\bf Main Results:} We evaluate the impact of different reverberation algorithms on auditory distance perception in virtual environments. Our main focus is to evaluate the distance compression characteristics of approximate vs. accurate interactive reverberation algorithms.   The objective is to determine if increase in physical accuracy leads to an increase in perceptual differentiation. 
In terms of approximate reverberation, we use a parametric Schroeder filter~\cite{Valimaki:2012:FYA:2713201.2713267} and perform dynamic calibration to improve its performance in terms of reverberation effects.
The accurate propagation effects are computed using a fast acoustic ray-tracer that can perform $50-100$ orders of specular and diffuse reflections at interactive rates to compute dynamic late reverberation on a multi-core PC~\cite{schissler2016interactive}. As the source or the receiver moves, the reverberation is recomputed and used for auralization.

A key issue in the use of parametric filter is tuning of various parameters and that can change the reverberation. In order to tune these filters for distance perception, we present a dynamic calibration algorithm that uses the $RT_{60}$ (reverberation time) and $DRR$ (direct-to-reverberant ratio) parameters. These two parameters are computed  from the impulse response based on the accurate ray tracing algorithm.
The Schroeder filter was appropriately scaled and spliced on the early part of the impulse response, starting at the approximate onset time of reverberation to match the $RT_{60}$ and $DRR$ of the ray-traced impulse response. In this way, our calibration algorithm tends to compute the best tuned filter for auditory distance perception.

We conducted two  separate studies to test auditory distance perception in virtual environments. The first study (static scene) is used to evaluate the effects of two reverberation methods in two rectangular-shaped rooms of different sizes for sources that were close to the listener, i.e., between $1m-5m$. The source and listener were at static locations. The second study (dynamic scenes) studied the effects of two reverberation methods in a large rectangular-shaped room for sources that were far away from the listener, i.e. between $10m-40m$. Furthermore, the listener in this study was moving in the environment, and thereby the reverberation effects perceived by the listener change. We analyze the results from these studies and observed the following results in VR:

\begin{itemize}

\item Auditory distance perception shows substantial compression in virtual environments and can be well-represented using a power-function.  Our power functions are similar to those reported by Zahorik~\cite{zahorik2002assessing}, though the exponents are different.

\item
  In case of a static scene, a \textit{well-tuned or calibrated} parametric reverberation filter can exhibit similar compression and perceived distance as the ray-traced reverberation method. The filter parameters, $RT_{60}$ and $DRR$, were matched to ray-traced generated impulse response parameters. This study indicates that in case of a static environment, only the $DRR$ is sufficient for a consistent and similar result. 
  
\item In case of dynamic scenes, a well-tuned or calibrated parametric reverberation filter shows similar compression characteristics as the accurate ray tracing method. However, the perceived distance is consistently higher for the ray tracing method. 
This finding indicates that matching the $DRR$s of two method is not sufficient for  distance perception in dynamic scenes. Instead, the full impulse response for the given source and listener positions also plays  an important role in terms acoustic distance estimation. This result suggests that accurate reverberation methods can lead to better reproduction of acoustic distances. 

\end{itemize}

\noindent The rest of the paper is organized as follows. In Section 2, we give a brief overview of prior work in acoustic simulation and auditory distance perception.  In Section 3, we discuss the calibration method used to match the filter to the impulse response. We describe our study for the static scene in Section 4 and present the experiment for the dynamic scene in Section 5.  Section 6 analyzes the results.

\section{Related Work and Background}
In this section, we give a brief overview of sound propagation, including the computation of impulse responses and reverberation. We also review many techniques for generating reverberation based on precomputed filters, geometric propagation and wave-based propagation.


\subsection{Acoustic Impulse Response and Reverberation}
The acoustic impulse response  filter refers to how an environment behaves when provided with an impulse for a given source and listener position in the environment. In particular, it captures the acoustic signature of an environment and gives us an idea of how a source would sound in that particular source-listener position configuration. In the real-world, acoustic impulse response measurements are widely used to characterize the acoustic characterization of an environment~\cite{Kuttruff:2007:BOOK}.
The impulse response can be used to study the acoustic characteristics of an environment and derive important acoustic metrics such as $RT_{60}$ (reverberation time), DRR (direct-to-reverberant energy ratio), $C_{80}$ (clarity), $D_{50}$ (definition), G (strength), Onset (onset delay), dir (onset direction), etc.


\vspace*{0.1in}

\noindent{\bf Reverberation:} Reverberation is the acoustic phenomenon of repeated sound reflections in an environment. 
Reverberation lends large environments a characteristic impression of spaciousness, and plays a crucial role in the acoustic design of buildings, concert halls, etc.  The two key characteristics for measuring this property are reverberation time ($RT_{60}$) and the direct-to-reverberant ratio ($DRR$).

$RT_{60}$ is defined as the time it takes for the sound in the environment to decay by 60 dB. $RT_{60}$ is one of the most important parameters in room and architectural acoustics and provides an approximation of  how long the reverberation effects will last in a given environment. For example, the recommended reverberation time for a classroom is $0.4-0.6$ seconds and is about $1.8-2.2$ seconds for a concert hall. Empirically, $RT_{60}$ is computed using the well-known Sabine's equation:
\begin{equation}
RT_{60} \approx 0.1611sm^{-1} \frac{V}{Sa},
\label{eqn:sabine_equation}
\end{equation}
where $V$ is the volume of the space, $S$ is the total surface area, and $a$ is the average absorption coefficient of the surfaces. This formulation doesn't require the exact impulse response of the environment and  is mostly used to estimate to the actual $RT_{60}$ in simple, rectangular rooms. The more accurate way of computing the $RT_{60}$ is based on reverse cumulative trapezoidal integration to compute the decay of the impulse response and then use a linear least-squares fit to compute the slope of the decay from $0$ dB to $-60$ dB~\cite{Kuttruff:2007:BOOK}.

The DRR is another well-known metric related to reverberation, and is formulated based on the energy ratio  between the energy of the sound coming directly from the source and its reflections.  Given the acoustic impulse response, the DRR can be estimated from the impulse responses by evaluating the onset and decay characteristics. Mathematically, DRR can be expressed as:
\begin{equation}
DRR = 10log_{10}(\frac{X(T_{0}-C:T_{0}+C)^{2}}{X(T_{0}+C+1:T_{n})^2}),
\label{eqn:drr_equation}
\end{equation}
where $X$ is the approximated integral of the impulse response, $T_{0}$ is the time of the direct impulse, $T_{n}$ is the time of the last impulse, and $C=2.5ms$ ~\cite{zahorik2002direct}.
Some recent work on human hearing has predicted that DRR can also provide absolute distance information, especially in reverberant environments~\cite{zahorik2002direct}.

\subsection{Sound Propagation}
At a broad level, sound propagation deals with modeling how  sound waves reflect, scatter, and diffract around obstacles as they travel through an environment. These techniques are used to measure the impulse response, which is composed of the direct sound, early reflections, and reverberation. The early reflections correspond to the sound reaching the listener after a small number of reflections and are used to localize the sound source(s). Sound propagation has been an active area of research in computational acoustics and virtual environments for many decades. At a broad level, prior   propagation methods can be classified into: use of parametric filters, geometric propagation, and wave-based propagation.

\subsubsection{Parametric Filters} 
Some of the simplest algorithms to model reverberation are based on parametric or artificial reverberation filters~\cite{Jot:AES:1991} that capture the statistics of reverberant decay using a small set of parameters.  The three main categories are delay networks, convolution-based algorithms, and physical room models~\cite{Valimaki:2012:FYA:2713201.2713267}. These are widely used in games and virtual environments because of their low runtime overhead. Typically the designer or artists divide a large scene into different reverberation zones or regions and compute separate  filters for each region. Moreover, different filters are interpolated at runtime to generate smooth acoustic responses. Many techniques have been proposed to automatically compute the reverberation parameters~\cite{Bailey:PATENT:2010,antani2013aural}.


\subsubsection{Geometric Acoustic Methods} 
These algorithms are  based on ray tracing and its variants (e.g., beam tracing or frustum) and assume that the sound travels along linear rays. Different techniques have been proposed to compute specular and diffuse reflections~\cite{Krokstad1968,Allen1979,Vorlander89}. They work well for high frequency sources, though some approximate techniques have also been proposed to approximate low-frequency effects such as edge diffraction. 
Over the last decade, many faster algorithms for interactive ray tracing have been proposed to accelerate sound propagation and these methods can exploit the parallel capabilities of current multi-core CPUs and GPUs. Earlier, interactive ray tracing algorithms were limited to compute only the early reflections~\cite{Funkhouser1998,10.1109/TVCG.2012.27} in dynamic scenes, while late reverberation was estimated using statistical approximation or precomputation methods~\cite{Antani:TOG:2012}. 
Recently, interactive techniques have been proposed to compute higher order specular and diffuse reflections using ray tracing~\cite{schissler2014high,schissler2016interactive}. These algorithms are able to compute $50-100$ orders of reflections in dynamic scenes at interactive rates on a multi-core desktop PC by exploiting the coherence of the sound field and performing backward ray tracing. This enables us to accurately compute early reflections as well as dynamic late reverberation effects. In this paper, one of our goals is to evaluate the  psycho-acoustic characteristics of these dynamic late reverberation algorithms for auditory distance perception.

\subsubsection{Wave-based Simulation}
  The wave-based algorithms numerically solve the acoustic wave equation and compute the sound pressure field and impulse responses.
  Recently, many  precomputation techniques have been proposed for interactive
applications \cite{Tsingos2007,Nikunj10,Mehra12a,Yero13}.  As compared to geometric methods, these algorithms are able to accurately model low frequency effects, but the time complexity increases as the fourth power of the frequency. As a result, they are only practical for low to medium range frequencies (e.g., less than $1$ or $2$ KHz). Furthermore, they are limited to static scenes due to the high precomputation overhead.

\subsection{Acoustic Distance Perception} 
There is considerable work in perception and VR literature on distance perception based on visual and/or aural cues~\cite{Finnegan:2016:CDC:2858036.2858065}. This includes work in VR on spatial perception with visual displays, audiovisual environments, and auditory displays~\cite{Meng:2006:DEV:1140491.1140523,sahm2005throwing,rebillat2012audio,interrante2006distance}. Interestingly, distance compression was observed in all such virtual environments. 
Other work includes the study of auditory perception in the presence of visual information~\cite{calcagno2012role}.

In terms of acoustic cues, most of the earlier research was on directional aspects of auditory localization~\cite{zahorik2002assessing}. Over the last few decades, there is considerable work on auditory distance perception as well as use of virtual acoustic techniques~\cite{bronkhorst1999auditory}. This includes evaluation of how well can humans estimate the stationary sound sources. Most of these studies suggest that listeners systematically underestimate distances to faraway sound sources.


\section{Calibrating Filter Parameters}
In this section, we present the details of the Schroeder filter used to generate reverberation effects. We also describe our dynamic calibration algorithm adjusts some parameters of this filter.

\subsection{Schroeder Filter} 
\begin{figure}[h]
\centering
\includegraphics[width=0.95\columnwidth]{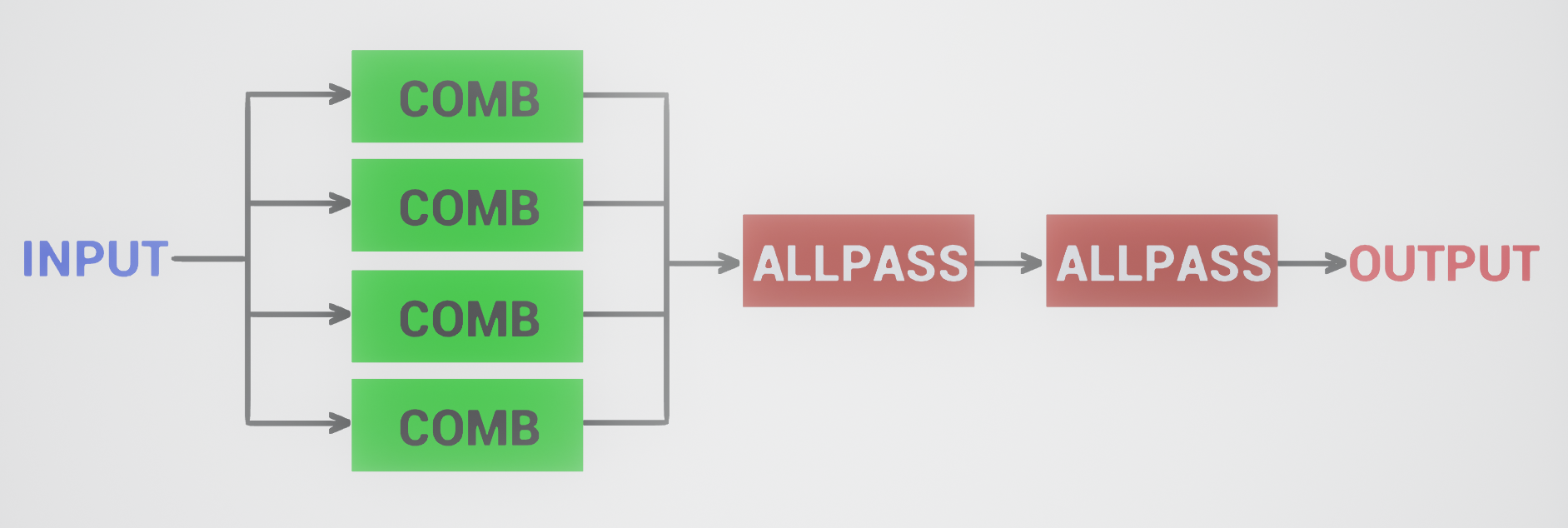}
\vspace*{-0.5em}
\caption{Schroeder reverberation filter: We highlight various components used in the design of such a filter. The comb filters produce reflections of the input sound, while the all-pass filters control the reverberator's wet and dry mix.}
\label{fig:schroeder}
\end{figure}

We use the Schroeder filter as an approximation method to compute reverberation. This filter is one of the oldest, most commonly used digital reverberation algorithms. It is easy to set up and customize to produce plausible reverberation characteristics in an environment.
An artificial reverberator must have a flat frequency response and  should be able to produce the desired density of reflections. The Schroeder filter uses a parallel bank of comb filters connected to a series of all-pass filters, as shown in Figure ~\ref{fig:schroeder}. The comb filters generate a repeated version of the input signals, while the all-pass filters keeps the frequency gain of the input as constant. The parallel-bank of comb filters combined with a series of all-pass filters  serves as a means to control the final output sound with a greater resolution. The overall specification of the Schroeder filter is controlled by different parameters, such as:   

\noindent {\em Room size}: This parameter specifies size of the room for which reverberation is being generated. Larger rooms tend to have longer reverb times.

\noindent {\em Pre-delay}: This parameter specifies the time it takes sound to reflect once.

\noindent {\em Damping}: It specifies the amount of attenuation applied to the high-frequency content of the sound. 

\noindent {\em Wet Gain}: This parameter specifies the amount of energy contained by the reverberation relative to the direct sound. 

\noindent {\em Dry Gain}: It specifies the amount of energy contained by the direction sound relative to the reverberation. 

\subsection{Calibration of Impulse Responses} 
\begin{figure}[h]
\centering
\includegraphics[width=0.95\columnwidth]{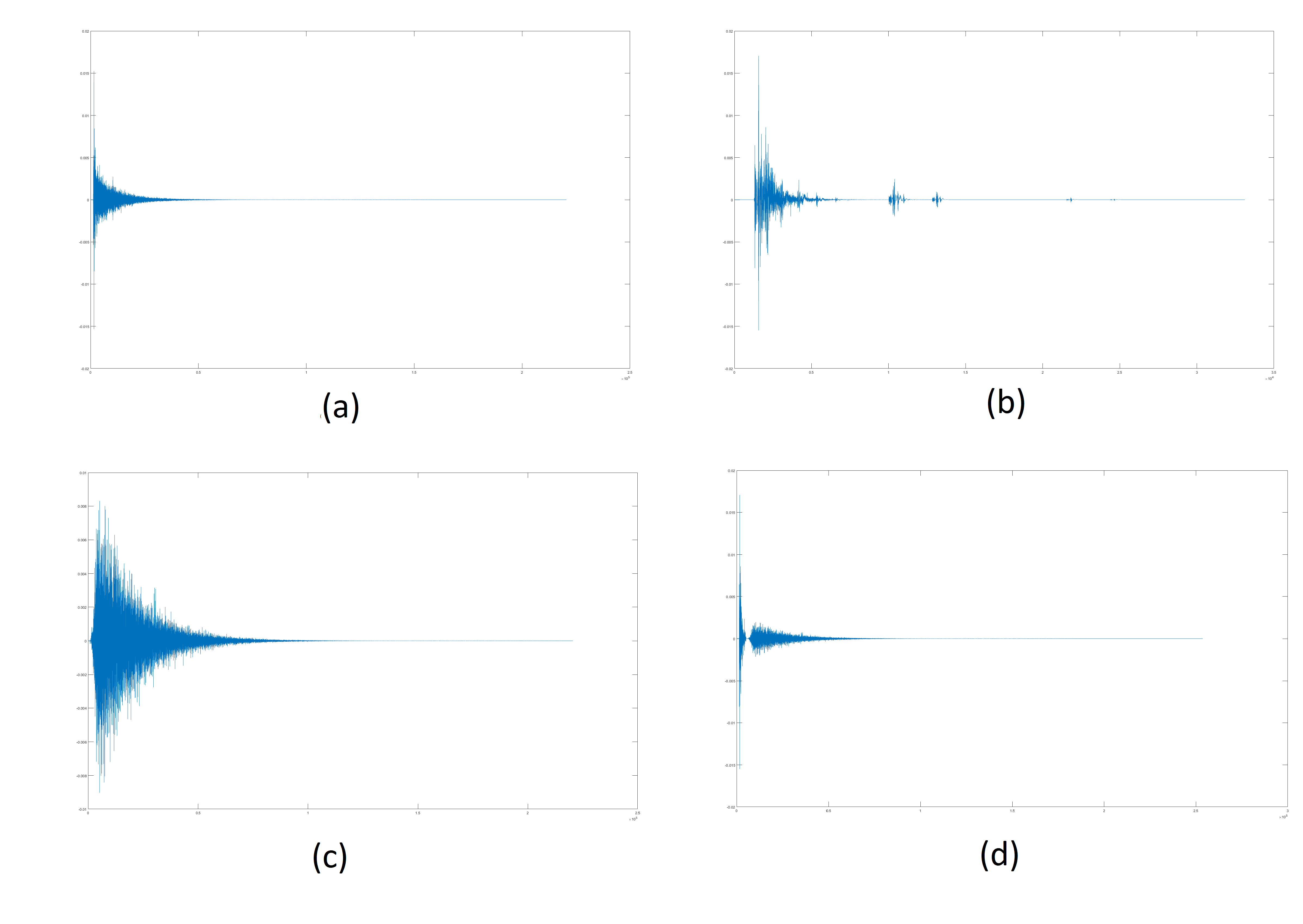}
\vspace*{-0.5em}
\caption{ (a) The full impulse response computed using the ray-tracer; (b) The early part of the impulse response computed using the ray tracer (using a different time scale); (c) The output of a generic Schroeder-reverberator; (d) The calibrated Schroeder filter with matching $RT_{60}$ and $DRR$ values as those of (a)}
\label{fig:allIRs}
\end{figure}

The performance and reverberation effects computed in an environment are governed by the different parameters used in the Schroeder filter formation. Since we are comparing two methods of reverberation, we need to make sure that early part of the impulse response for the two methods is exactly the same in our experiments. As a result, two versions of the impulse responses were computed for each source, listener position - the first one is the full impulse response computed using the ray tracer~\cite{schissler2016interactive} and the second one without the reverberant tail that contains only the early part of the impulse response as shown in Figure ~\ref{fig:allIRs}(a) \& (b).  We compute the $RT_{60}$ and DRR using the full impulse response, and adjust the parameters of the Schroeder filter till these parameters match.

\begin{algorithm}[h]
\caption{Impulse Response Calibration}\label{IRC}
\begin{algorithmic}[1]
\Procedure{IRC}{$FullIR,EarlyIR$}
\State $[RT_{60}^{Full}, DRR^{Full}]\gets GetParameters(FullIR)$ / FullIR is computed using ray tracing
\State $MaxNumberOfIterations \gets M$
\State $OnsetTimes \gets t_{firstsample} + [t_{1}..t_{n}]$
\State $Scaling \gets [s_{1}..s_{n}]$
\State $RF \gets SchroederFilter()$
\While{$i\leq MaxNumberOfIterations$}
\For {time in OnsetTimes}
\For {scale in Scaling}
\State $SyntheticIR \gets EarlyIR(time) + RF * scale$
\State $[RT_{60}^{Synthetic}, DRR^{Synthetic}]\gets GetParameters(SyntheticIR)$
\EndFor
\EndFor
\If {$RT_{60}^{Synthetic} - RT_{60}^{Full} \leq \epsilon_{RT_{60}}$ and $DRR^{Synthetic} - DRR^{Full} \leq \epsilon_{DRR}$}
\State $MatchedIR \gets SyntheticIR$ / i.e. $RT_{60}$ and DRR match with those of ray tracer
\EndIf
\State $i\gets i + 1$
\EndWhile\label{IRCendwhile}
\State \textbf{return} $MatchedIR$
\EndProcedure
\end{algorithmic}
\label{alg:calibration}
\end{algorithm}

Algorithm~\ref{alg:calibration} gives the details of calibration scheme.
$SchroederFilter$ is a generic Schroeder reverberation generator that produces a series of repeated, decaying impulses. The $GetParameters$ function computes the $RT_{60}$ and $DRR$, as described in the previous section. The $OnsetTimes$ defines a linear space of reverberation onset times. Typically, late reverberation starts around $t_{firstsample} = 80ms$~\cite{hidaka2007new} from the first impulse, so this space defines a time domain from $t_{firstsample}$ + $t_{1}$ to $t_{firstsample}$ + $t_{n}$. This linear space is used to determine the time at which the synthetic (Schroeder filter) reverberation is added to the early part of the impulse response. The $Scaling$, on the other hand, determines the space of multipliers that scales the synthetic reverberation computation. For each pair $\{time, scale\}$, the synthesized impulse response is constructed and its $RT_{60}$ and $DRR$ are evaluated. If these values match the $RT_{60}$ and $DRR$ of the full impulse response, based on the just-noticeable difference (JND) of the respective parameters($\epsilon_{RT_{60}}$, $\epsilon_{DRR}$), it corresponds to a considered a calibrated, matched impulse response.

\begin{figure}[h]
\centering
\includegraphics[width=0.95\columnwidth]{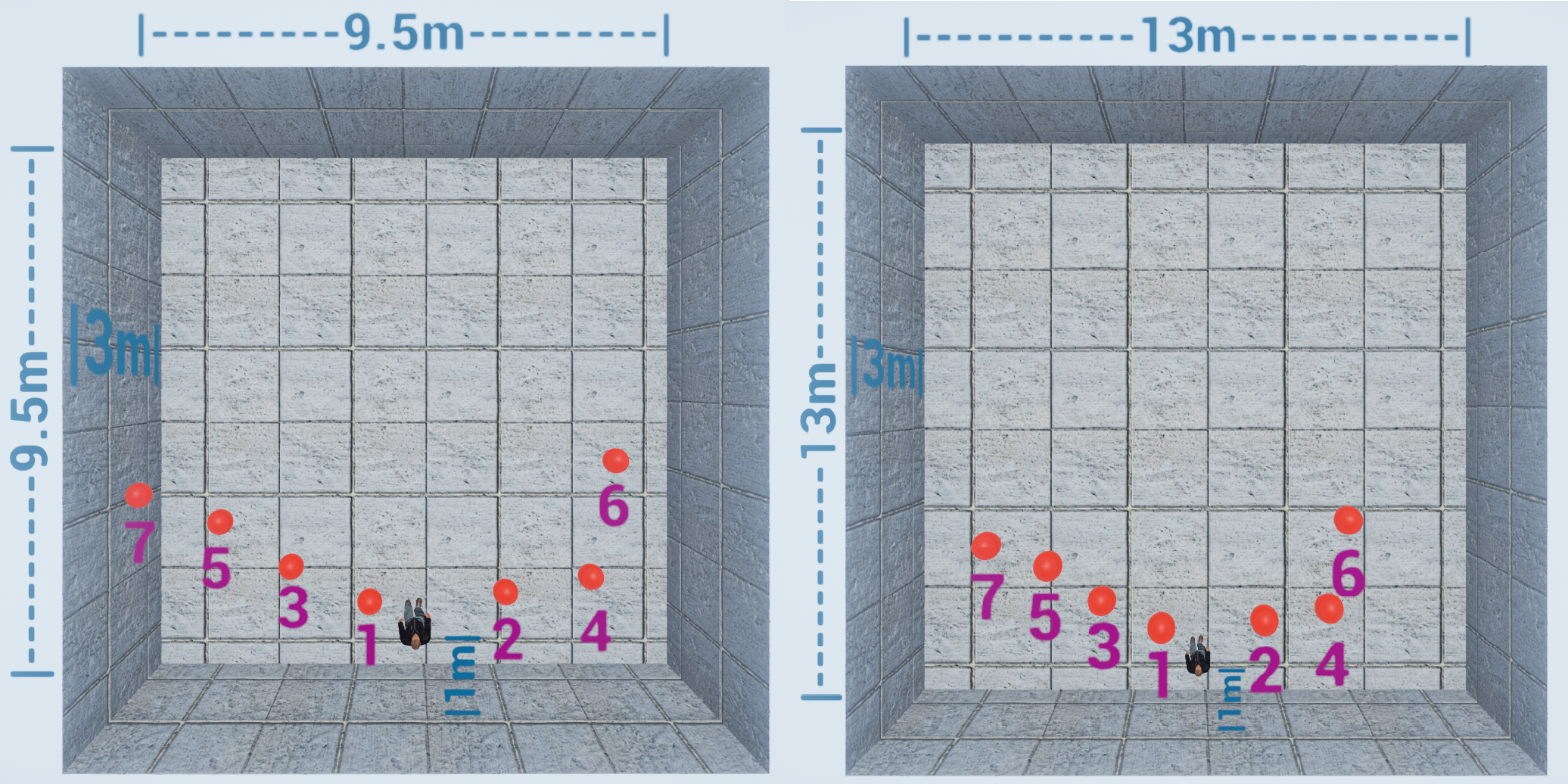}
\vspace*{-0.5em}
\caption{The two rooms used for the experiments. The red dots show the 7 randomly generated azimuths where the sources were kept. The numbers 1..7 indicate the distances at uniformly sampled points between [1m-5m] from the listener for both rooms. The subjects were placed 1m away from the wall.}
\label{fig:static_scene}
\end{figure}

\section{Experiment 1: Static Listener in a Reverberant Environment}

In this section, we give an overview of the static scene used in our evaluation. This study is based on a static sound source and a static listener.

\subsection{Participants} Twelve subjects took part in the study with informed consent. The ages ranged from $20$ to $31$ (mean = 25.2 and SD = 2.7). It include one female and eleven males.
All subjects reported normal hearing. 

\subsection{Apparatus} The set up consisted of a Dell T7600 workstation and the sound was delivered via a pair of  Beyerdynamic DT990 PRO headphones. The subjects were blindfolded for the study. The software to compute the $RT_{60}$ and $DRR$ is based on open-source MATLAB code ~\cite{IR_Stats}. The calibration and auralization were performed using in-house software, also written in MATLAB. 

\subsection{Stimuli} The stimuli consisted of a pre-recorded sound of a human male voice. The voice was played in two highly-reverberant virtual rooms of dimensions $9.5m \times 9.5m \times 3m$ and $13m \times 13m \times 3m$. The subjects were placed $1m$ away from one of the walls of both the rooms, as shown in Fig ~\ref{fig:static_scene}. The listener's head-orientation was fixed to look in the forward direction. Seven omnidirectional sound sources were placed in the azimuthal plane of the listener at distances of $1.0m, 1.7m, 2.3m, 3m, 3.7m, 4.3m$, and $5.0m$ at a constant height of $1.7m$. This was performed assuming the standard listener's height of $1.7m$ in the virtual environment.  The azimuthal angles for these $7$ source positions were pre-selected randomly: 3 positions between [-20$^{\circ}$ -45$^{\circ}$] and 4 positions between [20$^{\circ}$ 45$^{\circ}$] relative to the zero straight ahead, and kept the same throughout the study for all the subjects. Sound was rendered by constructing two impulse responses per source using the two reverberation methods - Schroeder reverberation filter and interactive ray tracing using the calibration scheme described above. The impulse responses were then convolved with the sound signal for auralization. The source sound power was $78$dB.

\subsection{Design \& Procedure} The environments were created to emulate a highly-reverberant room, similar to what a painted indoor room with no windows would sound.  The calibration algorithm described above was used to match the $RT_{60}$ and $DRR$ of the two reverberation methods, i.e., the Schroeder filter and interactive ray tracing. This was a within-subject study and all of the subjects experienced the acoustic effects from both methods of reverberation for all seven sources in the two rooms. The subjects were blindfolded, and the sound was delivered through headphones. This way we can accurately evaluate the acoustic effects.  In the absence of variable head-orientation, a generic HRTF-filter-based spatialization was used, as the evaluations only involved sources in the azimuthal plane. All the distance estimates were in meters. 

Before starting the experiment, the subjects were given an idea of the size of meter by showing them a meter-long strip of tape. After that, they were blindfolded and four sound clips were played: $1m$ away at 30$^{\circ}$ to the left, $1m$ away at 30$^{\circ}$ to the right, $5m$ away at 30$^{\circ}$ to the left, and $5m$ away at 30$^{\circ}$ to the right.  The subjects  were familiarized that these are the kind of sounds they should expect, without giving them the actual distances associated with these clips. The clips were rendered using the interactive ray tracing algorithm. The subjects were asked to rate the distances in meters on a scale of $1m-10m$.  The same metric was used in the subsequent experimental trials.

This was a within-subject study, and each subject rated the complete set of $7$ source positions within two rooms. The order of the sound files was randomized before starting a block of trials, and three such blocks were conducted, giving a total of $42$ evaluations per subject ($7$ source positions $\times$ $2$ rooms $\times$ $3$ blocks). The subject's head orientation was kept fixed, and the sound clip for each source position was played in a loop, until the listener responses. The subjects were allowed to take breaks. The experiment took an average of fifteen minutes per subject. No fatigue was reported. 


\begin{figure}[h]
\centering
\includegraphics[width=0.95\columnwidth]{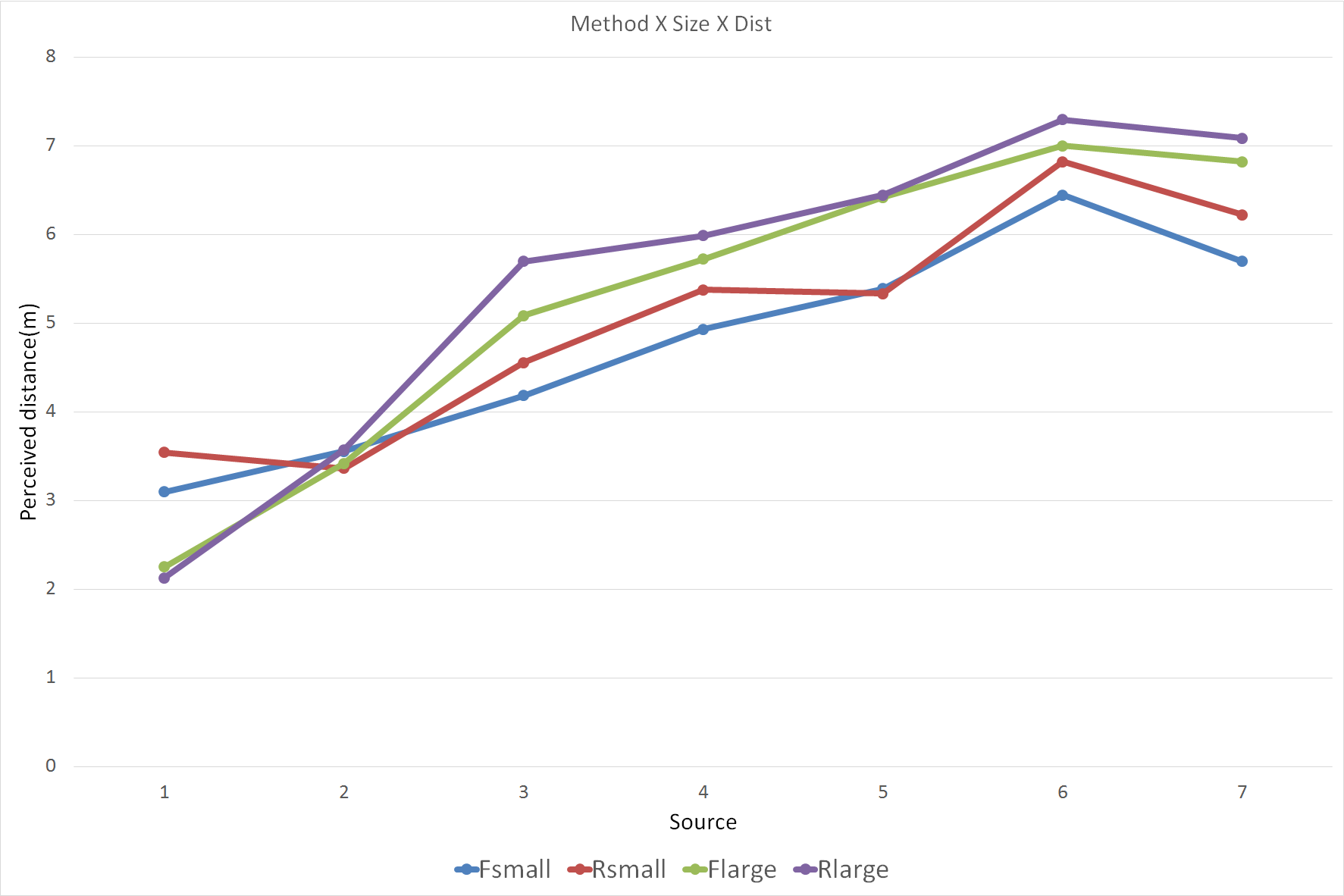}
\vspace*{-0.5em}
\caption{The average values as judged by the subjects in the two rooms for the two reverberation methods in the static scene. Fsmall (blue), Rsmall (red), Flarge (green), Rlarge (purple) represent small room with filter, small room with ray tracing, large room with filter, and large room with ray-tracing, respectively. The effect of the reverberation method (approximate vs. accurate) is null on the perception of distance, which is to be expected in a static scene where listeners would tend to use just the $DRR$ and intensity for the distance estimation. Due to our calibration routine, we ensured that the DRR of the Schroeder filter matched that of the interactive ray tracing method. These results suggest that auditory distance perception in static scenes is mainly governed by the DRRs.}
\label{fig:sizexdist}
\end{figure}

\subsection{Results} A 4-way ANOVA on block, distance, room, and reverberation method shows significant effects for distance ($F(6,66) = 56.46$, p $<$ $0.01$) and room($F(1,11) = 12.52$, p $<$ $0.01$), but none for the reverberation method or block. The two-way interaction between distance and room also shows significance ($F(6,66) = 11.23, p$ $<$ $0.01$).  This indicates that the room size has an effect on the perception of distance despite the distances being the same in the two rooms (Figure ~\ref{fig:sizexdist}). On the other hand, the  reverberation method  doesn't seem to affect the distance estimate. This can be explained if we take into account the fact that distance perception is a function of the $DRR$ and since the $DRRs$ were matched closely, in a static scene, the distance perception is similar. Moreover, the bigger room also has higher $DRR$ values compared to the smaller one.  The two reverberation methods did not noticeably affect the perception of distance in either of the rooms.  There is a significant two-way interaction between method and block, however, $F(2,22) = 4.4, p = .024$.  The pattern of this effect is modestly greater distance estimate in Block 1 for the ray tracing algorithm (mean $5.6 m$) relative to the Schroeder filter (mean $4.9 m$).  As this effect was absent in subsequent blocks,  the effect size was small. No further block effects were observed.

The results indicate that our method is sensitive to both distance and moderating effects (i.e., room size) for a static environment with sources relatively near to the listener. The absence of consistent effects of reverberation method over the present distances, together with the indication that larger rooms would increase perceived distance, motivated us to perform a second experiment with two changes:  First, the room size was increased to a much larger value, $45m$, allowing an expanded range of simulated distances.  Second, dynamic cues to auditory distance perception, based on listener's movement, were added.   The purpose was to determine if accurate reverberation  would expand perceived distance under these conditions.


\begin{figure}[h]
\centering
\includegraphics[width=0.95\columnwidth]{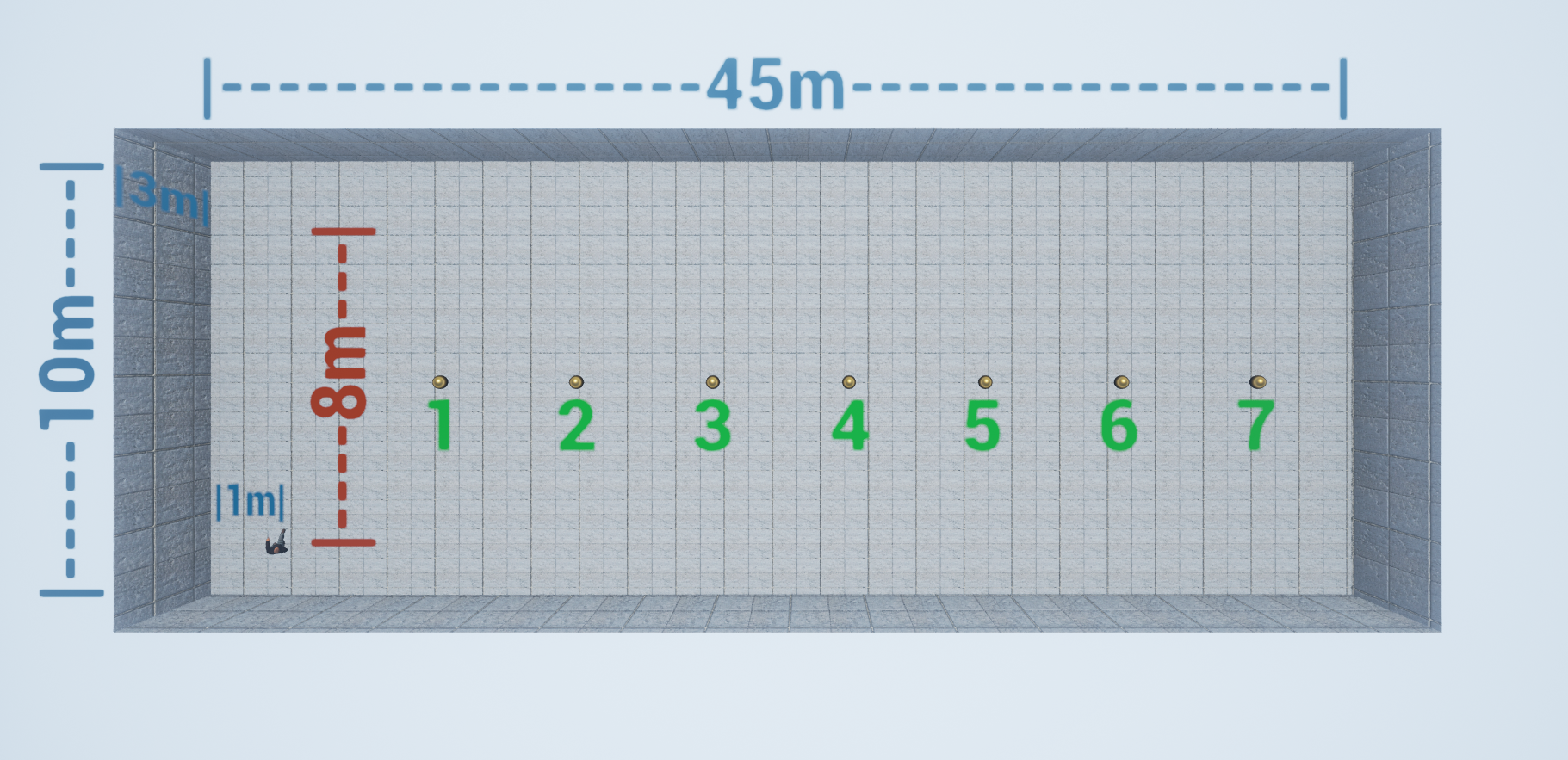}
\vspace*{-0.5em}
\caption{The room used for the experiment. The path marked in red is the walking path along which the subject walks. The sound sources are perpendicular to the walking path kept at increasing distances from it. The labels 1-7 show the different distances sampled uniformly from the range [10-40]m}
\label{fig:dynreverb}
\end{figure}

\section{Experiment 2: Moving Listener in a Reverberant Environment}

In this section, we present the study results in the dynamic scene. In this scenario, the listener moves and the movement results in dynamic reverberation effects.

\subsection{Participants} Seventeen participants took part in the study with informed consent. Their ages ranged from $19$ to $47$ (mean = $25.9$ and SD = $7.4$. Four females, thirteen males). The participants were recruited from the students and staff at the university. All participants reported normal hearing. 
\subsection{Apparatus} The set up consisted of a Dell T7600 workstation and the sound was delivered via a pair of  Beyerdynamic DT990 PRO headphones. The subjects were blindfolded for the study. The software to compute the $RT_{60}$ and $DRR$ was  based on open-source MATLAB code ~\cite{IR_Stats}. The calibration and auralization were done using in-house software, also written in MATLAB. 

\subsection{Stimuli} The source was a pre-recorded, broadband sound of human clapping. The virtual environment consisted of a rectangular room $45m \times 10m \times 3m$ with highly reflective walls to create a highly-reverberant environment with an eight meter walking path as shown in   Figure ~\ref{fig:dynreverb}. Seven omnidirectional sound sources were kept at increasing distances from the center of the path starting from $10m$ up to $40m$ in increments of $5m$. The sources were all kept at the same height of $1.7m$ from the floor. This value was chosen, assuming a standard listener height of $1.7m$ in the virtual environment. Sound was rendered by constructing two impulse responses per source and using the two reverberation methods - Schroeder filter and interactive ray tracing - and convolving them with the source signal. The source sound power was $78$dB. The details on how the impulse responses were generated are described below. 

\subsection{Design \& Procedure} A rectangular room was chosen as it provides the `best-case' scenario for the reverberation filter. The environment was created to emulate a highly-reverberant environment similar to a painted, concrete room with no windows. As mentioned in Section 2, the DRR is one of the main parameters in distance perception.
In order to make sure we're comparing the \textit{underlying methods} and not the specific parameters, we ensured that the  $RT_{60}$ and the $DRR$ are same for both reverberation methods.


\subsubsection{Training} Before the participants started the experiments, they completed a training task. The training part of the experiment took place in a real-world setting. An $8m$ long walking path was constructed and the sound sources were placed at $3m$ and $6$ from the center of the walking path, starting with $3m$. The participants were blindfolded before being led into the room so as to not give them an idea of the room dimensions. The dry (without reverberation) sound clip was played from a Harmon/Kardon HK $195$ desktop speaker. The participants were asked to point at the sound source with their right hand and keep pointing at the source as they walked along the $8m$ path. Since the participants were blindfolded, they were helped by the test administrators as they walked down the path. Once they reached the end of the path, the participants were asked to give their best evaluation (in meters) as to how far they thought the sound source to be when it seemed closest to them. The training task was then repeated with the source moved to $6m$. The subjects were told the actual distances at the end of the training. The training exercise was not meant to be an exact replica of the experiment as it was not possible to construct a physical room with the same kind of reverberance, as the one in the  virtual environment. Instead, the training was meant to give the participants a feeling as to what to expect and how to make judgements. 
Please refer to the supplementary video on how the training was performed.

\subsubsection{Method} This was a within-subject study and all the participants experienced both methods of reverberation. The walking in the virtual environment was not controlled by the participants; instead, the $81$ impulse responses per source were sampled such that each impulse response contributed to $0.1$m of the total $8$m for a human traveling at the average speed of $1.39$ $m/s$. The contributions from each of these $81$ impulse responses/sources were spliced  together (with interpolation) to create a sound file for each source. This sound file was played to the participants and they were asked to give the same estimate  as they performed in the training, i.e., the perceived distance (in meters) of the sound when it seemed to be the closest to them. The impulse responses were spatialized using a generic HRTF-filter. The participant's head orientation was fixed and they were always looking straight ahead.  Each participant rated the complete set of $7$ source positions $\times$ $2$ reverberation methods with the order of the sources randomized for each block, giving a total of $42$ ($7$ source positions $\times$ $2$ methods $\times$ $3$ blocks) judgements. The total time for the experiment, including the training, took around $15$ minutes. The participants were allowed to take breaks between blocks, as required. No fatigue was reported. 



\subsection{Results} A 3-way ANOVA on block, distance, and reverberation method shows the significant effects of method ($F(1,16) = 15.29$, p $<$ $0.01$) and distance ($F(6,96) = 29.12$, p $<$ $0.01$).  All two-way interactions (block-distance, block-method, distance-method) failed to show significance indicating that the reverberation method and distance effects are independent. This finding indicates that the shape of distance compression is statistically the same for both the reverberation methods and the ray tracing algorithm exhibits  an overall tendency to give longer distances. 

The null effect of block indicates that we should average over blocks in presenting data and that averaging does not obfuscate any trends in data - it only makes the data cleaner.

\section{Analysis} 

In this section, we analyze the results of our studied performed in the static and dynamic scenes. Acoustic distance perception is a complex phenomenon, and few studies have characterized its effect in virtual environments. Nevertheless, acoustic distance perception forms an important component of immersion in virtual environments and helps provide additional cues to the perception of the soundscape. 

\subsection{Data Fitting}

Zahorik~\cite{zahorik2002assessing} performed a comprehensive study of acoustic distance perception in virtual environments that assesses the  weights assigned to the principal cues. To analyze the data, \cite{zahorik2002assessing} fitted a  power function of the form:

\begin{equation}
D_{r} = kd_{r}^{a},
\label{eqn:powerfunction}
\end{equation}
where $D_{r}$ is the perceived distance, $k$ is a constant, $a$ is the power-function exponent that determines the function's rate of growth or decay, and $d_{r}$ is the actual source distance.  

Zahorik's data~\cite{zahorik2002assessing} found that perceived distance was a  power function  of the simulated distance with the power parameter averaging $0.39$.   A value less than $1.0$ means that perceived distance was highly compressive. The mutiplicative parameter was $1.3$, which would result in over-estimation of very low distance values.  There was also  substantial variability in the reported judgments across the individuals, particularly for distances $>$ $1m$. The tendency to compress perceived distance was consistent across the source signal type or  direction, but relative  weighting of  cues did vary with these factors. 

In our present data, the power function fit on the data generated the following functions:

\begin{equation}
D_{r}^{accurate} = 1.56{d_{r}^{0.58}},
\label{eqn:powerfunctionGSound}
\end{equation}

and

\begin{equation}
D_{r}^{filter} = 1.08{d_{r}^{0.66}}.
\label{eqn:powerfunctionFilter}
\end{equation}
The $R^2_{accurate}$ is $0.94$ while the $R^2_{filter}$ is $0.99$; thus both functions account well for the observed variability in the mean perceived distance.  The exponents exceed by $\sim$50\% the average value found by Zahorik~\cite{zahorik2002assessing} for a stationary listener.

\begin{figure}[h]
\centering
\includegraphics[width=0.95\columnwidth]{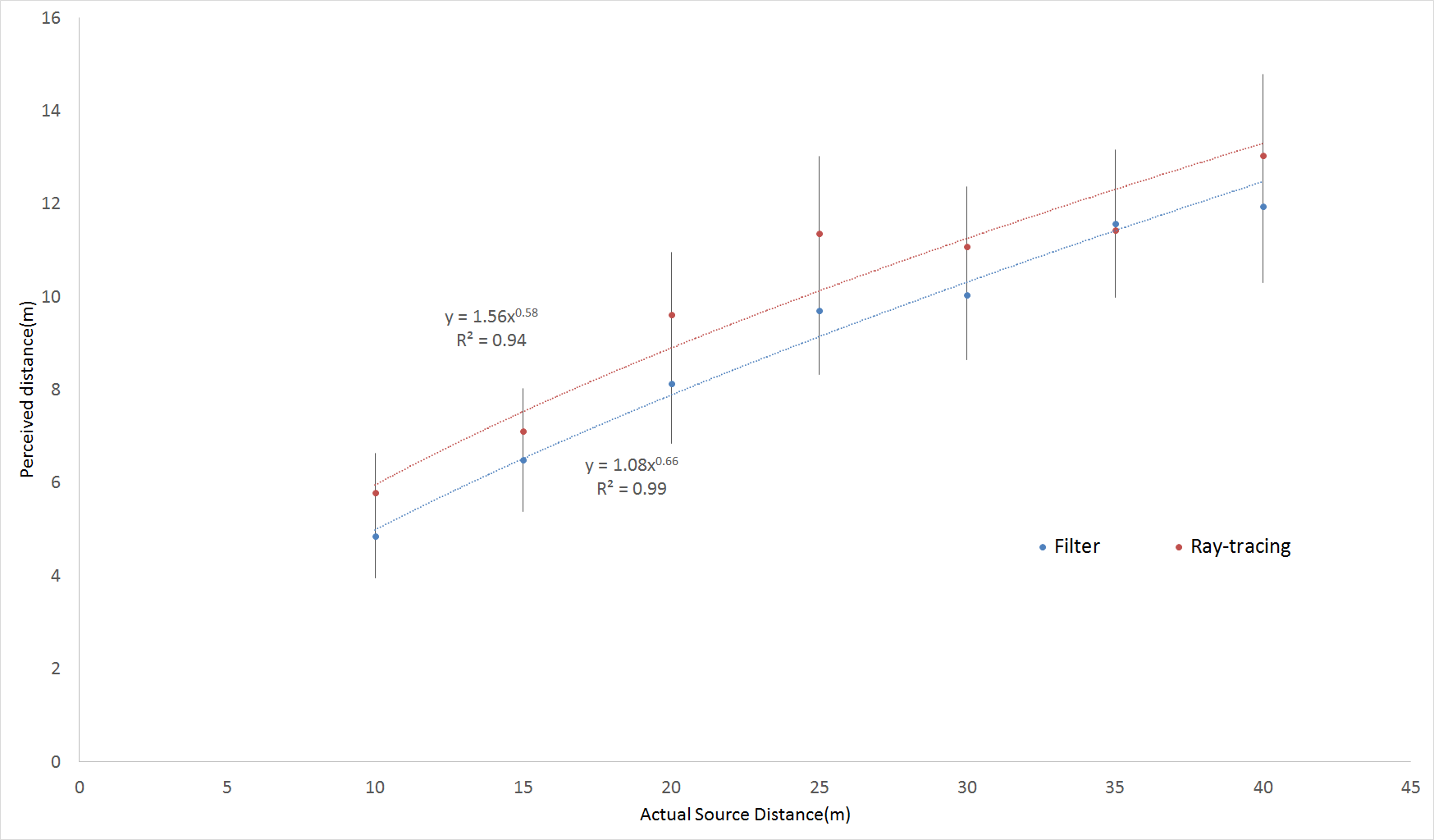}
\vspace*{-0.5em}
\caption{The power function fit of the distance data for the Schroeder filter (in blue, bottom curve) and interactive ray tracing (in red, top curve) algorithms.  This plot suggests that the compression  of  perceived  distance  relative  to  simulated  distance was comparable across both methods of generating reverberation. The distance perceived with accurate ray tracing algorithm exceeds that obtained with the filter method by essentially a constant amount.}
\label{fig:powerfit}
\end{figure}




The similarity of the power-function exponent for the two reverberation methods confirms the lack of interaction between physical distance and method in the ANOVA, which indicates that the compression of perceived distance relative to simulated distance was comparable across both methods of generating reverberation.  The effect of the greater multiplicative parameter for the accurate method is to move the responses closer to the true values for all distances measured in our study and any measured beyond this range of $10$m -- $40$m.  Any over-estimation resulting from a multiplicative parameter $>$ 1.0 would be expected for much smaller distances. The statistically confirmed result is that across the range of values examined here, the distance perceived with accurate ray tracing algorithm exceeds that obtained with the filter method by essentially a constant amount.   

The degree of compression observed here with virtual sound must be evaluated relative to the compressive perception found in reverberatory environments with real sound.  If we take the data from ~\cite{klatzky2003encoding} to provide a standard, linear compression of $0.7$ is to be expected with verbal report.  The linear fits to the present data were reasonable for filter and accurate (values of $R^2_{filter}$ and $R^2_{accurate}$, respectively), and the corresponding slopes were $0.24$ and $0.23$.  In this context, we can estimate the additional compression due to simulation as a multiplicative factor on the order of $\frac{1}{3}$ giving us a good idea what to expect in terms of perceived distance when using virtual acoustics.

\subsection{Discussion}
The results in the static scene seem to suggest that matching the
DRRs of the approximate method based on Schroeder filter to that
of ray tracing method may provide the same level of accuracy in
terms distance perception. However, in case of dynamic scenes that is not sufficient and the accurate ray-tracing algorithm results in
better distance estimates. Interestingly, in most applications, the
parameters of the Schroeder filter are tuned by an artist or by using
some presets for a set of different environments. It is not clear
whether those methods can ensure that the DRRs of the Schroeder
filter would match those computed using the accurate ray-tracing
method. Our Algorithm 1 provides an automatic mechanism to
compute Schroeder filter with little manual efforts. Moreover, the
reverberation filters are mostly designed for rectangular rooms, and
are based on Sabine’s Equation. It is not clear whether they will
provide the same level of reverberation effects or distance estimates
for non-rectangular rooms.
\section{Conclusions, Limitations, and Future Work}

In this paper, we studied the impact of different sound propagation
algorithms on auditory distance perception. In particular, we compared
the performance of approximate techniques based on parametric
filters with accurate techniques based on interactive ray tracing
in static and dynamic scenes. Our study shows that although the
compression characteristics of two methods are similar, the more
accurate propagation method results in less distance compression
in VR, especially in dynamics scenes with a moving listener. This
finding suggests that accurate reverberation effects in a VR system
can be perceptually useful for different applications.
This work has some limitations. The geometric acoustic methods
are accurate at high frequencies, and it would be useful to combine
them with wave-based methods for low frequencies. Besides
sound intensity and DRR, there are many cues other that may affect
auditory distance perception, including the at-the-ear spectrum
due to the sound absorbing properties of the air and binaural differences
~\cite{zahorik2002direct}, and it would be useful to evaluate them as part of
future work. We would also like to extend our evaluation to environments
with moving sound sources, dynamic obstacles, nonrectangular
scenes, and varying the level of reverberation. It would
be useful to combine our results with other cues (e.g., visual perception).
Ultimately, we hope to develop good VR systems with
multi-modal capabilities (including sound), where researchers from
other fields (e.g., psychology) can evaluate different hypothesis.

\bibliographystyle{abbrv}
\bibliography{main}
\end{document}